\documentclass[twocolumn,showpacs,amsmath,amssymb]{revtex4}

\usepackage{epsfig}

\usepackage{dcolumn}

\usepackage{bm}

\usepackage{amsmath}

\usepackage{amsfonts}

\usepackage{amssymb}

\usepackage{graphicx}%

\newcommand{\beq}{\begin{equation}}
\newcommand{\eeq}{\end{equation}}
\newcommand{\beqn}{\begin{eqnarray}}
\newcommand{\eeqn}{\end{eqnarray}}

\newcommand{\slp}{\raise.15ex\hbox{$/$}\kern-.57em\hbox{$ \partial $}}
\newcommand{\lnA}{\raise.15ex\hbox{$/$}\kern-.57em\hbox{$A$}}
\newcommand{\lnC}{\raise.15ex\hbox{$/$}\kern-.57em\hbox{$C$}}

\begin{document}

\title{Non adiabatic features of electron pumping through a quantum dot in the 
Kondo regime}

\author{Liliana Arrachea}
\affiliation{Departamento de F\'{\i}sica de la Materia Condensada and BIFI, Universidad
de Zaragoza, Pedro Cerbuna 12, 50009 Zaragoza}

\author{Alfredo Levy Yeyati and Alvaro Martin-Rodero}
\affiliation{Departamento de F\'{\i}sica de la Materia Condensada, 
Universidad Aut\'onoma de Madrid, E-28048, Madrid, Spain.}

\date{ }
\begin{abstract}
We investigate the behavior of the dc electronic current $J^{dc}$ in
an interacting quantum dot driven by two ac local potentials oscillating
with a frequency $\Omega_0$ and a phase-lag $\varphi$. We provide analytical
functions to describe the fingerprints of the Coulomb interaction in an 
experimental
$J^{dc} \; \mbox{vs} \; \varphi $ characteristic curve. We show that  
the Kondo resonance reduces
at low temperatures
the frequency range for the linear behavior of $J^{dc}$ in $\Omega_0$ to take
place and determines the evolution of the dc-current as the temperature increases.
\end{abstract}

\maketitle

\section{Introduction}
In the beginning of the new century we are witnessing an increasing interest
towards dc transport induced by pure ac fields. Quantum pumps, where transport
is generated by applying harmonically time-dependent gates oscillating with a 
phase-lag $\varphi$ at the walls of semiconducting quantum dots, are 
paradigmatic examples realized in the laboratory \cite{switkes,pumpex,pepper}.
On the other hand, the possibility of exploring the Kondo regime in 
semiconducting and carbon nanotube quantum dots provides 
a unique test system to understand the role of electronic correlations 
in quantum transport \cite{kondodot1,kondodot2}. The combination of ac 
pumping mechanisms with many-body interactions constitutes a challenging 
avenue of research. On the experimental side these studies are likely to 
be feasible in the near future, since although these setups employ very 
slowly oscillating fields, great efforts are currently being devoted to 
increase the range of operational frequencies \cite{pepper}. The use of 
superconducting junctions as ac generators seems to be a promising 
methodology in this direction \cite{supjunc}. 

Since the celebrated proposals of Refs. \cite{butad,brouwer}, the 
``adiabatic approximations''
are at the heart of the theoretical work on pumping in quantum dots driven 
at their walls
 \cite{brou2001,mobu,otros}. 
Within these approximations the induced dc-current 
$J^{dc}$ is proportional to the
pumping frequency $\Omega_0$, describing the regime
where $\Omega_0 << \tau^{-1}$, $\tau$ being the characteristic time for 
the electrons to travel through the dot.
Few theoretical studies have 
addressed the problem of many-body interactions in quantum pumps
\cite{konpump,slavebos,das}. Most of the
work has been centered in ``adiabatic 
approximations'' \cite{konpump,slavebos}, and the electronic 
interactions are usually included within the slave boson mean-field 
approximation \cite{slavebos}, 
which does not properly account for  inelastic many-body effects.
 
In the present work, we also focus on near-equilibrium regimes, 
where $\Omega_0$ is lower
than  the Kondo temperature $T_K$, but we explore the effect of the interactions
beyond the ``adiabatic'' regime.
To this end we combine two methods: (i)  
the treatment of the interactions by a second-order self-energy as considered  in 
Ref.\cite{aa},  which has been a successful
tool to study dc transport in the Kondo regime \cite{aa,self},
and (ii) the Keldysh 
Green's functions formalism with the Fourier representation of Ref.
\cite{lilipum}, which has been used to study models of
non-interacting quantum pumps at arbitrary
frequency \cite{lilipum,lilipum1}.
We provide analytical expressions to identify the fingerprints of the interactions 
in the $J^{dc} \; \mbox{vs} \; \varphi$
characteristic curve, which is the feature usually explored experimentally \cite{switkes}.
We anticipate that the development of Kondo resonance imposes limits in the range of 
frequencies
for the $J^{dc} \propto \Omega_0$ behavior to be observed while 
inelastic scattering induced 
by the interactions tends to restore such a behavior.

The work is organized as follows. In the next section we present the model and technical
details on the theoretical approach. We present results in section IV. Section V is devoted
to a summary and conclusions.

\section{Theoretical formulation}
\subsection{Model}
\begin{figure}
{\includegraphics[height=40mm,
  keepaspectratio,
clip]{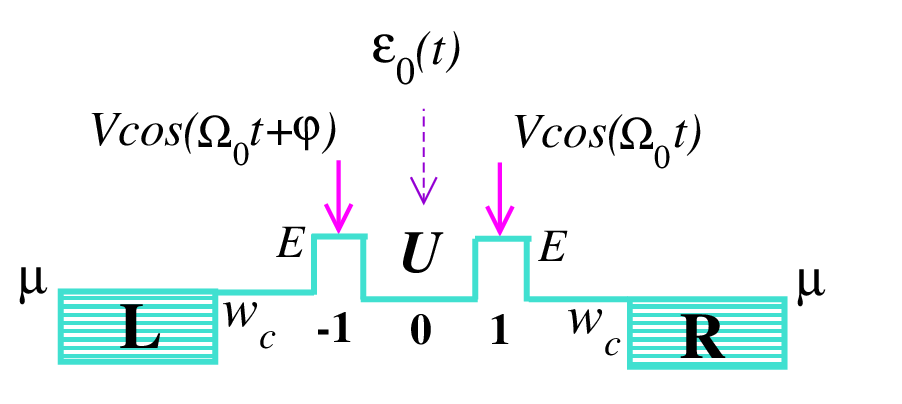}} \caption{(Color online) Sketch of
the setup. The two external ac potentials, and  the  induced
 potential at the interacting site are indicated with solid and dashed arrows, 
respectively.}
\label{fig1}
\end{figure}
  
We describe the quantum pump in terms of a generalized Anderson impurity model
(see sketch of Fig. \ref{fig1}):
\begin{eqnarray}\label{ham}
H(t)& &=\sum_{\alpha=L,R,k_{\alpha},\sigma}
\varepsilon_{k_{\alpha}} n_{k_{\alpha},\sigma}- 
w_c \sum_{\alpha, k_{\alpha},\sigma}( c^{\dagger}_{k_\alpha,\sigma } c_{l_{\alpha},\sigma}
 + H. c.) \nonumber \\
& & - w \sum_{l=-1,\sigma}^0 
(c^{\dagger}_{l,\sigma} c_{l+1,\sigma }+ H.c)
\nonumber \\
& &+ \sum_{l=-1,\sigma}^1 \varepsilon_l (t) c^{\dagger}_{l,\sigma}
 c_{l,\sigma } +U n_{0,\uparrow}n_{0,\downarrow},
\end{eqnarray}
being $\varepsilon_l (t)=\delta_{l,-1}[E+V\cos(\Omega_0 t + \varphi)]+
\delta_{l,1}[E+V\cos(\Omega_0 t )]$. The dot with Coulomb interaction $U$ is
inserted between two
barriers of height $E$ at which two ac fields are applied with amplitude
$V$. The degrees of freedom of the reservoirs are denoted with $k_{\alpha}$,
being $\alpha=L,R$, which are at equal chemical 
potential 
$\mu$ and temperature $T=1/\beta$.
The reservoirs are attached at the positions $l_{\alpha}=\pm 1$ of the 
central structure through hopping terms with amplitude $w_c$.

\subsection{Charge currents}
Following the procedure of
Refs. \cite{lilipum}, the dc-current along the structure can be expressed as
a sum of an {\em elastic} and an {\em inelastic} component as follows:
\begin{equation}\label{jdc}
J^{dc}  = J^{el}+J^{in}= \sum_{k=-\infty}^{+\infty} \int_{-\infty}^{\infty}
\frac{d \omega}{2 \pi} [ I^{el}(k,\omega)+ I^{in}(k,\omega)],
\end{equation}
where, adopting units with $e=\hbar=1$, and, to simplify, omitting  explicit reference
 to the spin index:
\begin{eqnarray}\label{jel}
I^{el}(k,\omega) &=& 2 \Big[f(\omega-k\Omega_0)-f(\omega)\Big] 
\Gamma(\omega) \Gamma(\omega-k\Omega_0)
\nonumber \\
& & 
\times \sum_{l_{\alpha}, l_{\alpha}'=\pm 1} l_{\alpha}
 |{\cal G}_{l_{\alpha}, l_{\alpha}'}(k,\omega-k\Omega_0)|^2,
\end{eqnarray}
where $\Gamma(\omega)= 2 \pi |w_c|^2 \sum_{k_{\alpha}}
\delta(\omega - \varepsilon_{k_{\alpha}})$ is the tunneling rate from the structure to 
the reservoirs and  ${\cal
G}_{l,l'}(k,\omega)$ is  the $k$-th Fourier coefficient 
of the retarded Green's function:
\begin{equation}
G^R_{l,l'}(t,t') = \sum_{k=-\infty}^{+\infty}  e^{-i k\Omega_0 t} \int_{-\infty}^{+\infty}
\frac{d\omega}{2 \pi} e^{-i \omega (t-t')}{\cal G}_{l,l'}(k,\omega),  \nonumber \\
\end{equation}

The {\em elastic} component takes into account processes where 
electrons propagate coherently along the dot structure experimenting virtual
 absorption
or emission of quanta of frequency $\Omega_0$ at the pumping centers. 
 The ensuing expression coincides with the 
non-interacting one, except for the fact that the retarded Green's function corresponds
in the present case to the interacting system described by the full Hamiltonian 
(\ref{ham}). Instead,
 the  {\em inelastic} contribution accounts for processes in which 
electrons experiment decoherence
originated in the many-body interactions.  
 It reads:
\begin{eqnarray}\label{jin}
I^{in}(k,\omega) & = & 
2 \sum_{k'=-\infty}^{+\infty} \mbox{Re} \Big\{ \Gamma(\omega)
\Big[\lambda^>(\omega) \Sigma^<_0(k',\omega-k\Omega_0)
\nonumber \\
& &-\lambda^<(\omega) \Sigma^>_0(k',\omega-k\Omega_0)\Big] \times \nonumber \\
& & 
\Big[\sum_{l_{\alpha}=\pm 1} l_{\alpha} 
{\cal G}_{l_{\alpha},0}(k-k',\omega-(k-k')\Omega_0)
\nonumber \\
& & \times 
{\cal G}^*_{l_{\alpha},0}(k,\omega-k\Omega_0)\Big] \Big\},
\end{eqnarray}
being $\lambda^{<}(\omega)=if(\omega)$ and  $\lambda^{>}(\omega)=-i(1-f(\omega))$,
which depend on the
Fermi function 
$f(\omega)=1/(e^{\beta(\omega - \mu)} + 1)$. 
We have also introduced the Fourier representation of Ref. \cite{lilipum} in the
lesser and bigger self-energies $\Sigma^{<,>}_0(k,\omega)$, which describe 
the many-body effects due to 
the Coulomb interaction:
\begin{equation}
\Sigma^{<,>}_{0}(t,t') = \sum_{k=-\infty}^{+\infty}  
e^{-i k\Omega_0 t} \int_{-\infty}^{+\infty}
\frac{d\omega}{2 \pi} e^{-i \omega (t-t')} \Sigma^{<,>}_{0}(k,\omega). \nonumber
\end{equation}

\section{Treatment of the Coulomb interaction}
\subsection{Time-dependent Hartree-Fock approximation}
While the above expressions are in principle exact, the many-body self-energies must be 
calculated
at some level of approximation. The lowest order in the interaction $U$ corresponds to the
time-dependent-Hartree-Fock (TDHF) approximation. The many-body problem is reduced to
consider Hamiltonian (\ref{ham}) with $U=0$ and 
a renormalized level given by
\begin{equation}
\varepsilon^{TDHF}_l(t)=
\varepsilon_l(t)+\delta_{l,0} 
U \sum_{k=-\infty}^{+\infty}  n^{TDHF}_{0,\sigma}(k) e^{-i k \Omega_0 t}.
\end{equation} 

The Fourier components of the local particle
 density $n^{TDHF}_{0,\sigma}(k)$ must be evaluated self-consistently. 
This level of perturbation theory leads to a vanishing inelastic contribution 
$I^{in}(k,\omega)$ and  it is not suitable 
for the description of Kondo physics. It is, however, interesting to notice that 
this simple approximation already introduces a
non-trivial ingredient in the pumping problem.
 Namely, the effective emergence of an additional
pumping center in the interacting site (see Fig. \ref{fig1}).

\subsection{Second order self-energy approximation}
In order to go beyond the TDHF description 
 we consider the second order self-energy (SOSE) given by
 the  bubble diagram of Ref. \cite{aa} and generalize the procedure of that work
to situations with an harmonic dependence on time. Concretely, we consider
the following
lesser and bigger components of the self-energy:
\begin{eqnarray}
& & \Sigma_0^{>,<}(k,t-t') = \pm i U^2 
\sum_{k_1,k_2=-\infty}^{+\infty} 
G_0^{0,>,<}(k_1,t-t') \nonumber \\
& & \times [G_0^{0,>,<}(k_2,t-t')]^* G_0^{0,>,<}(k-k_1+k_2,t-t'),
\end{eqnarray}
 where the propagators are $G_0^{0,>,<}(k,\omega)\equiv G_{0,0}^{0,>,<}(k,\omega)$ with
\begin{eqnarray}\label{gles0}
 & & G_{l,l'}^{0,>,<}(k,\omega)= \sum_{\alpha, k'} 
 {\cal G}^0_{l,l_{\alpha}}(k+k',\omega-k'\Omega_0)  \nonumber \\
& &
 \times  \lambda^{>,<}(\omega - k'\Omega_0)[{\cal G}^0_{l',l_{\alpha}}(k',\omega-k'\Omega_0)]^*, \\
& & G_{l,l'}^{0,>,<}(k,t-t') =  \int_{-\infty}^{+\infty}
\frac{d \omega}{2 \pi} e^{-i \omega (t-t')} G_{l,l'}^{0,>,<}(k,\omega), \nonumber 
\end{eqnarray}
being the {\em non-equilibrium} retarded Green's functions ${\cal G}^0_{l,l'}(k,\omega)$,
 the solution of the Dyson's equation corresponding to the Hamiltonian (\ref{ham}) with
$U=0$ and 
\begin{equation}
\varepsilon^{SOSE}_l(t)=
\varepsilon_l(t)+ \delta_{l,0}
 \sum_{k=-\infty}^{+\infty} \varepsilon_0^{eff}(k) e^{-i k \Omega_0 t}.
\end{equation}
 The Fourier components of the 
effective potential $\varepsilon_0^{eff}(k)$, are determined self-consistently from the 
condition that the occupation of the interacting site 
evaluated within the SOSE approximation equals the one evaluated with (\ref{gles}), i.e.
\begin{eqnarray} \label{self}
n^{SOSE}_{0 \sigma}(k) & \equiv & -i\int_{-\infty}^{+\infty} \frac{d\omega}{2\pi}
 G_0^{<}(k,\omega) \nonumber \\
& = & -i\int_{-\infty}^{+\infty} \frac{d\omega}{2\pi} G_0^{0,<}(k,\omega),
\end{eqnarray}
where the lesser Green's of the first equality contains the full dressing by  the
self-energy:
\begin{eqnarray}\label{gles}
 G_{l,l'}^{>,<}(k,\omega)& & = \sum_{\alpha, k'} 
 {\cal G}_{l,l_{\alpha}}(k+k',\omega-k'\Omega_0) \times \\
& &
  \lambda^{>,<}(\omega - k'\Omega_0)[{\cal G}_{l',l_{\alpha}}(k',\omega-k'\Omega_0)]^* +
 \nonumber \\
& &  \sum_{k', k''}   {\cal G}_{l,0}(k+k'-k'',\omega - (k'-k'')\Omega_0) \times \nonumber \\
& &
  \Sigma^{>,<}(k'',\omega - k'\Omega_0) [{\cal G}_{l',0}(k',\omega-k'\Omega_0)]^*,\nonumber
\end{eqnarray} 
while that of the second one corresponds to (\ref{gles0}).
The above procedure ensures the fulfillment of
 Friedel-Langreth sum rule in the equilibrium limit \cite{aa}.
Unfortunately, however, this procedure is not enough to ensure 
the conservation of the inelastic component of the current.
 
 The retarded self-energy is then obtained from
\begin{equation}
\Sigma^R_0(k,t-t')=\Theta(t-t')[\Sigma_0^{<}(k,t-t')-\Sigma_0^{>}(k,t-t')],
\end{equation}
 and it is introduced in the Dyson's equation:
for the  full dressed retarded Green's functions ${\cal G}_{l,l'}(k,\omega)$.
The latter are evaluated by using the renormalization procedure of 
Ref. \cite{lilipum}. The algorithm is combined with the use of fast Fourier transform
between the variables $t-t' \leftrightarrow \omega$. 

\subsection{Low amplitude and low frequencies expansion}
For small pumping amplitudes and frequencies, it is possible to derive an approximate
expression for the dc current that is accurate up to
 ${\cal O}(V^2)$ and 
up to ${\cal O}(\Omega_0)$.  Such a procedure would correspond to a generalized ``adiabatic
approximation'' within the present formalism and will allow us to get analytical expressions
to gain insight on the behavior of the dc-current. 
 
As a first step, we truncate the harmonics of the
induced potentials up to $\varepsilon_0^{eff}(0)$ and $\varepsilon_0^{eff}(1)=
 [\varepsilon_0^{eff}(-1)]^*=V_{eff}$, and the harmonics of the self-energy up
to $\Sigma(0,\omega)$ (higher harmonics involve terms ${\cal O}(V^2)$ and
${\cal O}(U^2 V)$). The Dyson's equation for the retarded Green's function reads:

\begin{eqnarray}
G^R_{l,l'}(t,\omega) & & \sim  G^0_{l,l'}(\omega) +  \\
& & \sum_{j=-1}^1 G^R_{l,j}(t,\omega+ \Omega_0) 
\varepsilon^{eff}_j(1) e^{-i\Omega_0 t} G^0_{j,l'}(\omega) + \nonumber \\
& & \sum_{j=-1}^1 G^R_{l,j}(t,\omega - \Omega_0) 
\varepsilon^{eff}_j(-1) e^{i\Omega_0 t} G^0_{j,l'}(\omega), \nonumber
\label{dyretap}
\end{eqnarray}
where the pumping potentials contain the external time-dependent fields as well as
the time-dependent potential
induced by the interactions:
\begin{equation} 
\varepsilon^{eff}_j(1)= V[\delta_{j,-1} e^{-i \varphi}+\delta_{j,1}] + 
V_{eff} \delta_{j,0},
\end{equation}
with $[\varepsilon^{eff}_j(1)]^*=\varepsilon^{eff}_j(-1)$.
 The 
Green's functions $G^0_{l,l'}(\omega)$ correspond to the stationary part of the 
Hamiltonian. The solution of (\ref{dyretap}) up to the first order in 
$\varepsilon^{eff}_j(k)$ casts
\begin{equation}
G^R_{l,l'}(t,\omega)= 
\sum_{k=-1}^1 {\cal G}_{l,l'}(0,\omega) e^{-i \omega_0 t},
\end{equation}
with
\begin{eqnarray}
{\cal G}_{l,l'}(0,\omega)& & =  G^0_{l,l'}(\omega) \nonumber \\
{\cal G}_{l,l'}(\pm 1 ,\omega)& & = \sum_{j=-1}^1 G^0_{l,j}(\omega \pm  \Omega_0) 
\varepsilon^{eff}_{j}(\pm 1)G^0_{j,l'}(\omega). \label{retp}
\end{eqnarray}
Eq. (\ref{self}) leads,   within this approximation,
to the following self-consistency condition to evaluate $V_{eff}$:
\begin{eqnarray}
V_{eff} &=& U \sum_{\alpha=L,R,j=-1}^1
\int_{-\infty}^{\infty}\frac{d \omega}{2\pi} \Gamma_0(\omega)
f(\omega) \varepsilon^{eff}_j (1) \nonumber \\
& & \times [ G^0_{0,j}(\omega+\Omega_0)
G^0_{j,l_{\alpha}}(\omega) G^0_{0,l_{\alpha}}(\omega)^* \nonumber \\
& & + G^0_{0,l_{\alpha}}(\omega)
G^0_{0,j}(\omega)^* G^0_{j,l_{\alpha}}(\omega - \Omega_0)^* ]. 
\end{eqnarray}
A rough estimate of the pumping potential {\em without self-consistency}
is:
\begin{equation}
\varepsilon_0^{(0)}(1)  \sim U V \lambda_0 ( 1 + e^{-i \varphi}),
\end{equation}
being
\begin{eqnarray}
\lambda_0 & = & \int_{-\infty}^{\infty} \frac{d \omega}{2\pi} \sum_{\alpha=L,R}
\int_{-\infty}^{\infty} \frac{d \omega}{2\pi} \Gamma_0(\omega)
f(\omega) \nonumber \\
& & \times [ G^0_{0,1}(\omega+ \Omega_0)
G^0_{1,j_{\alpha}}(\omega) G^0_{0,j_{\alpha}}(\omega)^* \nonumber \\
& & + G^0_{0,j_{\alpha}}(\omega)
G^0_{0,1}(\omega)^* G^0_{1,j_{\alpha}}(\omega - \Omega_0)^* ].
\end{eqnarray}
Thus, within this rough approximation, $V_{eff} \sim 
\varepsilon_0^{(0)}(1) = UV \lambda_0 [1 + \cos(\varphi) -i 
\sin(\varphi)]$. Within the self consistent procedure, however,
$V_{eff} = V^{'}_{eff} + i V^{''}_{eff}$, with both real and imaginary parts
of the form
$V^{' ('')}_{eff} \sim    A^{'('')}_0 
\sin(\varphi)+ A^{'('')}_1 (1+ \cos(\varphi))$.
On the other hand, expanding the above expressions up to the first order in 
$\Omega_0$ we find $V_{eff} \sim V_{eff}^0 + V_{eff}^1 \Omega_0$. 

Expanding (\ref{jel}) in powers of $\Omega_0$, casts for the first order contribution:
\begin{eqnarray}
J^{dc} & & \sim J^{el} \sim \Omega_0  \Gamma^2 (\mu) \nonumber \\
& & \sum_{k=-1}^1 k \{ |{\cal G}_{1,1}(k, \mu)|^2 - 
|{\cal G}_{-1,-1}(k, \mu)|^2 \nonumber \\
& & +  |{\cal G}_{1,-1}(k, \mu)|^2 - |{\cal G}_{-1,1}(k, \mu)|^2 \}.
\label{jelp}
\end{eqnarray}
Substituting the 0th order term in $\Omega_0$ of
 (\ref{retp}) into (\ref{jelp}), it is found
\begin{eqnarray}
J^{el}  & & \sim  \Omega_0  \Gamma^2 (\mu) \sum_{j \neq l} 
[ \varepsilon^{eff}_j(1)  \varepsilon^{eff}_l(1)^* - \nonumber \\
& & 
\varepsilon^{eff}_j(-1)  \varepsilon^{eff}_l(-1)^*]
 ( \lambda_{l j}-\lambda_{jl} ),
\end{eqnarray}
being 
\begin{eqnarray}\label{lij}
\lambda_{jl} &=& \sum_{l_{\alpha}=\pm 1}  l_{\alpha} G^0_{l_{\alpha},j}(\mu)
G^0_{l_{\alpha},l}(\mu)^* \gamma_{lj}(l_{\alpha}), \nonumber \\
\gamma_{lj}(-1) &=&  \sum_{l_{\alpha}=\pm 1}G^0_{j,l_{\alpha}}(\mu)
G^0_{l,l_{\alpha}}(\mu)^*, \nonumber \\
\gamma_{lj}(1) &=&  \sum_{l_{\alpha}=\pm 1}G^0_{j,l_{\alpha}}(\mu)
G^0_{l,-l_{\alpha}}(\mu)^*
\end{eqnarray}
In the case of the rough estimate for the induced pumping potential 
$\varepsilon_0^{(0)}(1)$ defined above, 
a relation between the current and the phase-lag of the form
$J^{el} \propto \sin (\varphi)$, as in the non-interacting case, is obtained. However, when
 the pumping potential is evaluated self-consistently, the current
behaves as follows:
\begin{eqnarray}
J^{dc} & & \sim J^{el} \sim \Omega_0 
\frac{(\Gamma^{el})^2}{(\Gamma_0)^4} V 
\Big[ V \sin(\varphi) \lambda^{el}_1 \nonumber \\
& & + \Big(V^{'}_0  \sin(\varphi)
(\cos(\varphi)+1)+ V^{''}_0 \sin^2(\varphi)\Big) \lambda^{el}_2  \Big],  \label{pertel}
\end{eqnarray}
where we have defined 
$\Gamma^{el}=\Gamma(\mu)$, being $\Gamma_0=4 |w|^2/\Gamma(\mu)$, which is
approximately
the effective tunneling rate from the interacting site to the 
leads. The dimensionless functions
$\lambda_i^{el}$ as well as $V^{'}_0, V^{''}_0$
depend on $V_{eff}$ and on the parameters $\lambda_{lj}$ of eq. (\ref{lij}).
 It is important to note that Eq. (\ref{pertel})
reduces to the the non-interacting result 
 $J^{dc} \sim V^2 \sin(\varphi)$
originally proposed in  Ref. \cite{brouwer},
 for $V_0^{'}=V_0^{''}=0$.
 Such a dependence on $\varphi$
has been observed experimentally 
 in Ref. \cite{switkes} and 
is a manifestation  of 
quantum interference between processes of coherent emission and
absorption of quanta at the two pumping centers. In the interacting
system the terms $\propto \lambda_2^{el}$ of Eq.(\ref{pertel}) are a 
consequence of additional
interference  with scattering events at 
the  pumping center induced by the interactions. 

The inelastic contribution exhibits the same structure, except for a
 different prefactor $\Gamma^{in}= -2 \mbox{Im}[\Sigma^{eq}_0(\mu)]$, 
being $\Sigma^{eq}_0(\omega) \sim \Sigma^R_0(0,\omega)$
the SOSE of the system with $V=0$, instead of $\Gamma^{el}$.
  For low $T$, $\Gamma^{in} \sim 
(\omega-\mu)^2 $, thus, for low $\Omega_0$
 $J^{in} \sim 0$ and the elastic processes account for
the full dc transport.

\section{Results}
\subsection{Weakly interacting regime}
\begin{figure}
{\includegraphics[width=80mm,
  keepaspectratio,
clip]{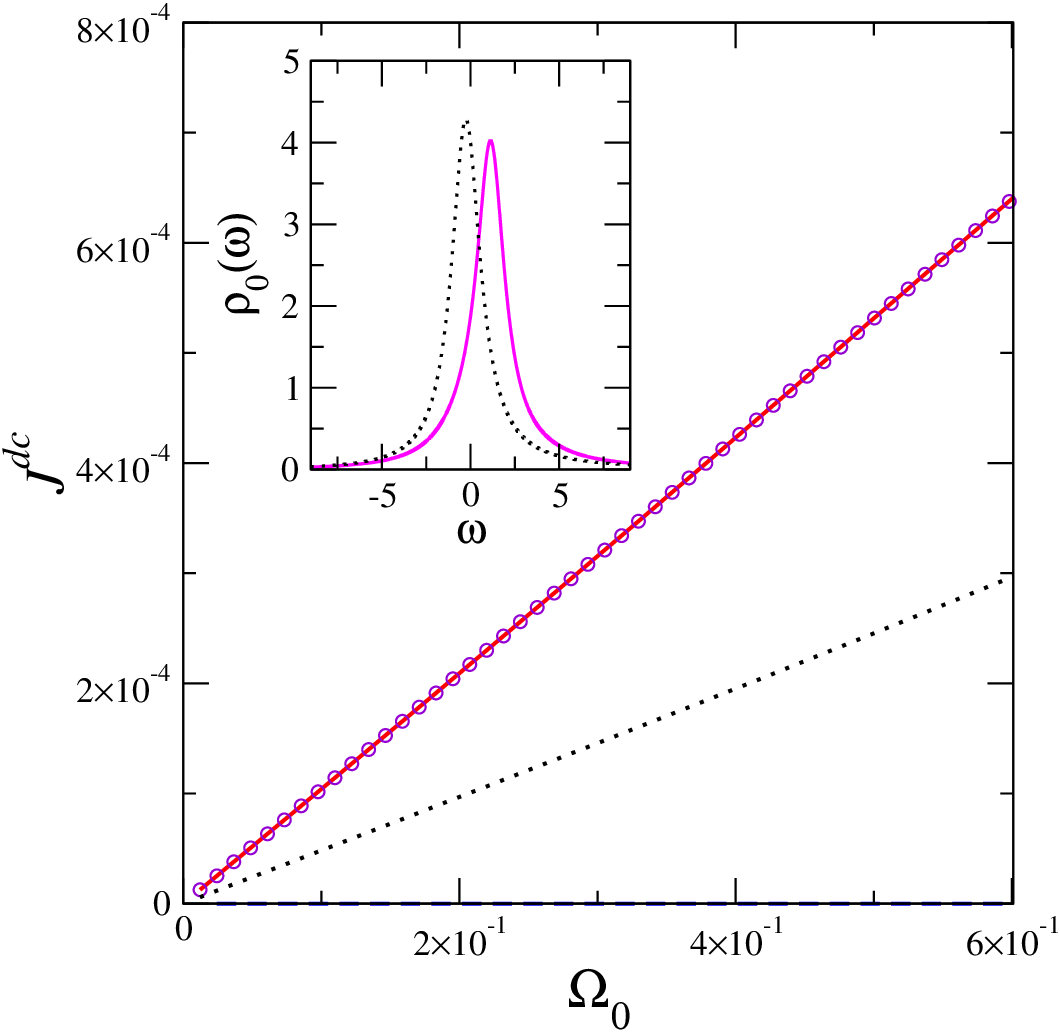}} \caption{(Color online) $J^{el}$ (solid line) , 
$J^{in}$ (dashed line)
and $J^{dc}$ (circles) for $\varphi=\pi/2$,
as functions
of the driving frequency $\Omega_0$ for $U=1$. The inelastic component is $J^{in}\sim 0$.
The dc-current for the non-interacting system ($U=0$) is also shown in dotted lines
 for comparison. The corresponding local densities of states at
 the interacting site $\rho_0(\omega)$ are indicated in the inset (solid lines), along
with the non-interacting one (dotted lines).Other parameters are $E=2,w=2,w_c=8$, $V=0.4$ and $\mu=2$. 
(Energies,  frequencies and currents are expressed in units of $\Gamma_0$).
}
\label{fig2}
\end{figure}
We begin with the
 analysis of the effect of the interactions in the behavior of $J_{dc}$
as a function of  $\Omega_0$  at  $T=0$ for a given chemical potential
$\mu$ and phase-lag $\varphi$. In   Fig. \ref{fig2} we present 
the behavior of the different contributions
to $J^{dc}$ obtained from the  numerical solution of the full
Dyson's equation retaining self-energy components $\Sigma_0^{>,<}(k,\omega)$
up to $|k|=2$, with the self-consistent evaluation of $\varepsilon_0^{eff}(0)$,
and $\varepsilon_0^{eff}(1) \equiv V_{eff}$.
Energies,  frequencies and currents are expressed in units of $\Gamma_0$.
 The inset shows
the corresponding local density of states at the interacting site 
$\rho_0(\omega)=-2 \mbox{Im}[{\cal G}_{0,0}(0,\omega)]$ along with the density
of states corresponding to the non-interacting system ($U=0$). The position of the
resonant peak experiments a shift equal to $\varepsilon_{eff}(0)$. 
 For low $U$
the effects of $\Sigma(k,\omega)$ are vanishingly small and the description 
effectively reduces to TDHF. The induced $V_{eff}$ is also very small 
($V_{eff} \sim 1\times 10^{-3} <<V$). 
For low $U$, the inelastic contribution to the current (${\cal O}(U^2)$)
is negligible. Therefore, within this regime,  $J^{dc}$ 
 is qualitatively similar to the non-interacting one, shown in dotted line.
As the hybridization between the dot and the side reservoirs is sizable, the resonant
peak is wide and the ``adiabatic'' ($\propto \Omega_0$ behavior) is observed within
 a wide range of $\Omega_0$.

\subsection{Kondo regime}
\begin{figure}
{\includegraphics[width=80mm,
  keepaspectratio,
clip]{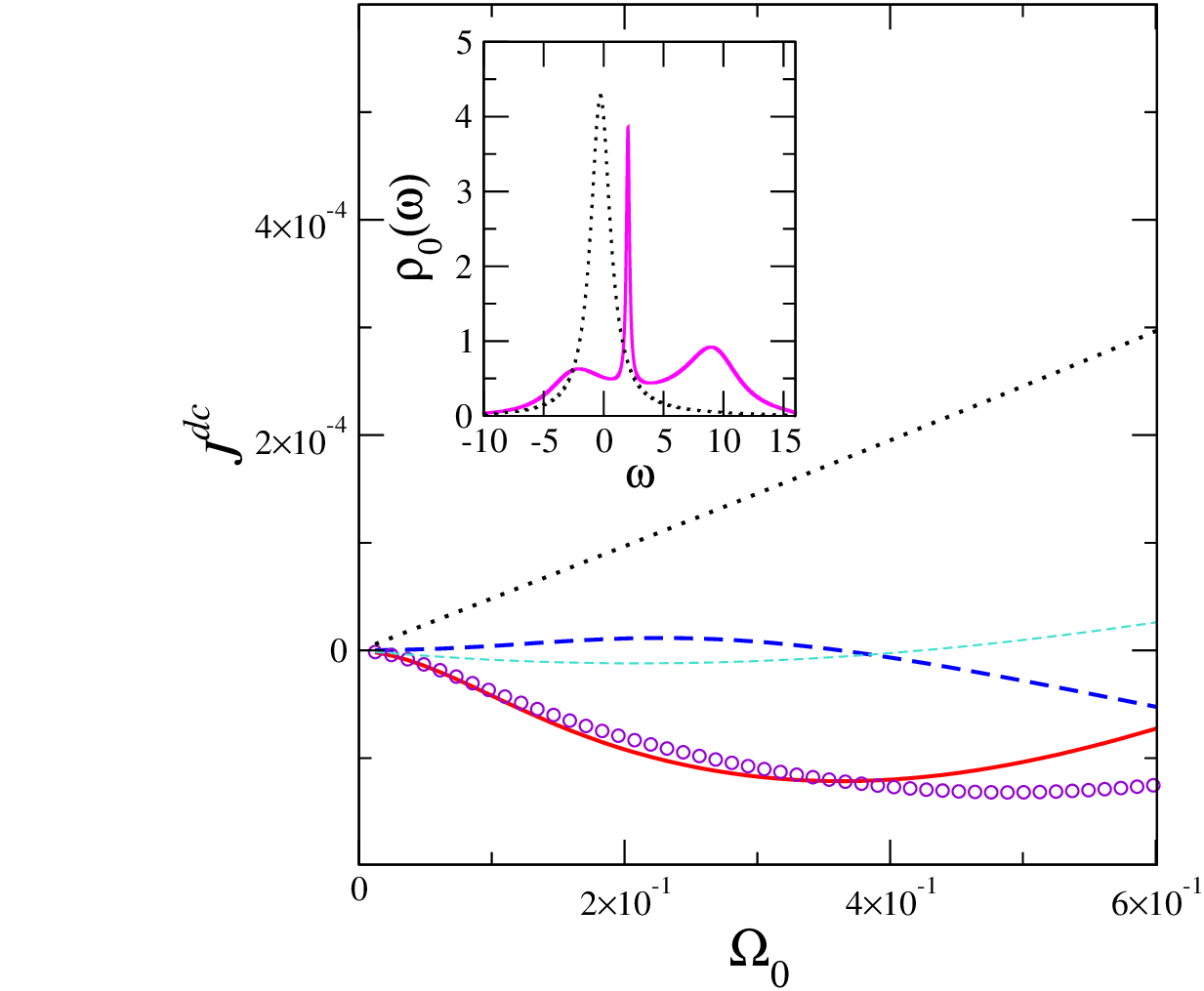}} \caption{(Color online) $J^{el}$ (solid line) , 
$J^{in}$ (dashed line)
and $J^{dc}$ (circles) for $\varphi=\pi/2$,
as functions
of the driving frequency $\Omega_0$ for $U=10$. The details are the same as in
Fig. \ref{fig2}.
}
\label{fig3}
\end{figure}
Let us now  analyze the more subtle Kondo regime, which takes place at higher $U$.
In Fig \ref{fig3} we show a series of plots similar to those of Fig. 
\ref{fig2} but corresponding to a value of $U$ for which the Kondo effect takes place.
For these parameters, the density of states at the interacting site $\rho_0(\omega)$
exhibits the characteristic Kondo resonance at $\omega=0$
 with the two high-energy side features centered at $\omega=\pm U/2$ (see inset).
As a function of $\Omega_0$ the behavior of $J^{dc}$ significantly departs from
the non-interacting one in this regime and there are several issues to comment in connection to 
Fig. \ref{fig3}. First, it is clear that self-energy effects now play an
important role. This is evident in the drastic changes experimented
by the density of states as well as in the fact that the current departs from the
behavior predicted by the TDHF approximation, which is shown in light dot-dashed line
in the figure. 

The second feature to remark is that, although inelastic processes
are negligible for low $\Omega_0$, as discussed in Section III.C, they become
sizable as  $\Omega_0$ increases even at $T=0$
(see dashed line of the main frame of  Fig. \ref{fig3}).

 The third remarkable issue is the lose of the ``adiabatic'' behavior, namely, the
departure from the linear dependence in $\Omega_0$. We identify two ingredients
that contribute to this effect: (i) The first one is the induced pumping 
potential at the interacting site, which becomes sizable  $V_{eff}>1 \times 10^{-2}$ and
changes significantly with $\Omega_0$. 
 This feature manifests itself even
at the TDHF level, in which case the range of pumping frequencies 
where $J^{dc (TDHF)} \propto \Omega_0$  becomes very narrow 
($\Omega_0 <2 \times 10^{-2}$). (ii) The other ingredient is 
the development of the Kondo resonance.
Recalling that the 
underlying 
assumption for low frequency
expansions like the one leading to Eq. (\ref{pertel})
is that the typical width of the energy levels of the structure 
is  $\tau^{-1} >>\Omega_0$, it can be understood that 
 the reduction of the width of the resonant level
due to the Kondo effect originates  a concomitant
reduction of
the range of $\Omega_0$ for the  ``adiabatic''
 behavior to be observed.
For the parameters of Fig. \ref{fig3}, which correspond to $T_K \sim 0.5$, 
such a linear behavior is not captured even close to the lowest pumping frequency 
considered ($\Omega_0= 1\times 10^{-2}$). 

\begin{figure}
{\includegraphics[width=80mm,
  keepaspectratio,
clip]{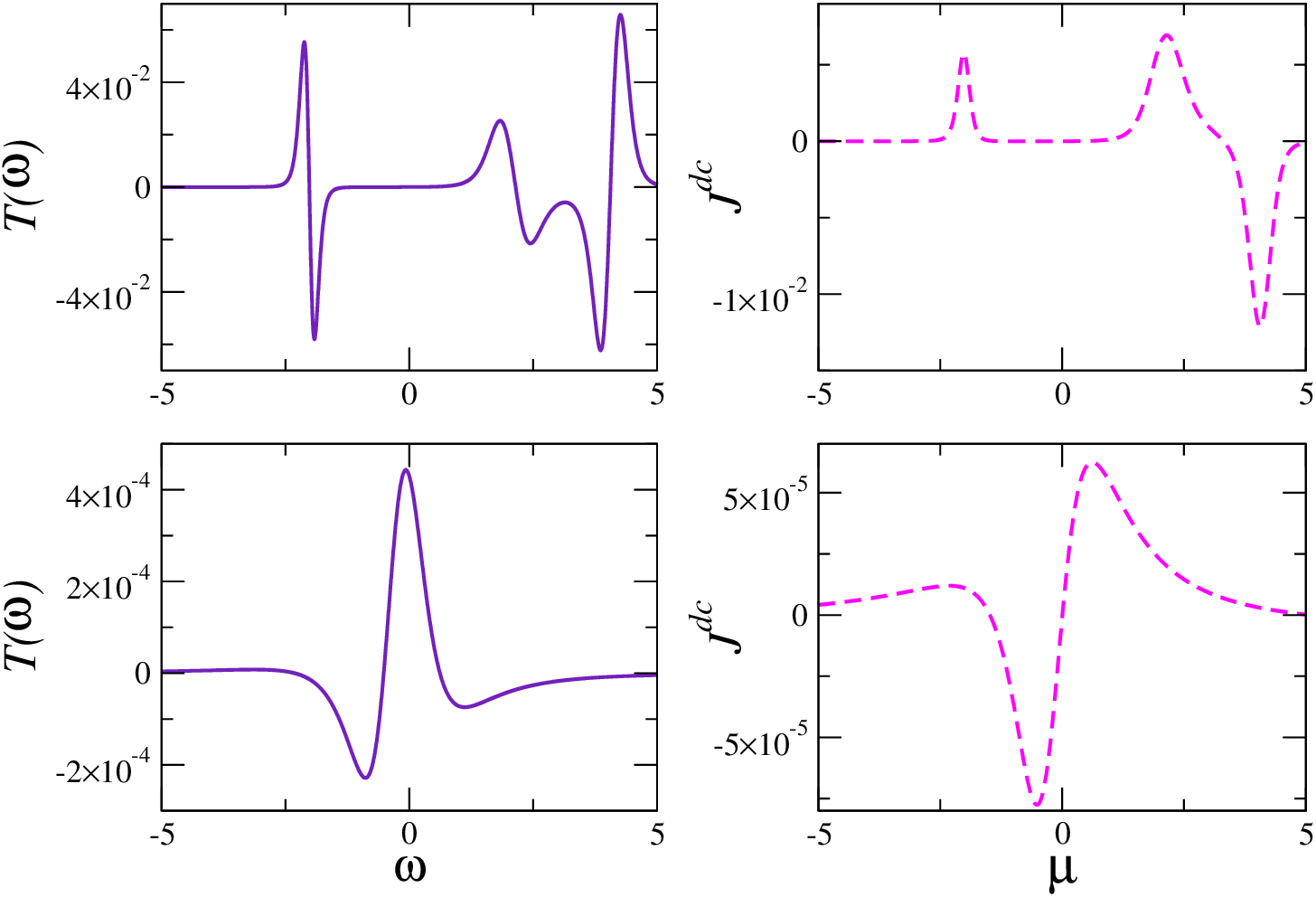}} \caption{(Color online) ``Transmission function''
 $T(\omega)$ for a noninteracting
structure with $\Omega_0=0.1$ and  $w_c=0.64$ (upper left panel) and $w_c=8$ (lower left panel).
 Other details are the same as in
Fig. \ref{fig2}.
}
\label{fig4}
\end{figure}

The final issue worth of mention is the inversion in the sign of the current
with respect to the non-interacting case. In general, even in the simpler case
of a system without many-body interactions,
the issue of the sign of the current is one of the most delicate
features to predict in
a problem of quantum pumping. In the case of structures with low hybridization with
 the contacts presenting a landscape of well separated resonances, the issue of the sign
of the current has been analyzed in detail in Refs. 
\onlinecite{mobu,lilipum,lilidir}. In those
works, the
origin for the sign inversion of the current has been identified to be 
 the interference between two resonant electronic levels mixed by  a
high pumping frequency. In order to gain insight on the different processes involved in the present 
problem,
let us first analyze the behavior of the sign of the current in
the non-interacting limit of our setup of Fig. \ref{fig1}. In that case the dc current 
(\ref{jdc}) can be also written as follows \cite{lilipum}:
\begin{equation}
J^{dc}_{non-int}=\int_{-\infty}^{+\infty} d\omega f(\omega) T(\omega),
\end{equation}
where the ``transmission'' function $T(\omega)$ depends on the Green's functions
of the system with $U=0$. Plots of the function $T(\omega)$ and its integral 
between $-\infty$ and $\mu$, which is equivalent to $J^{dc}_{non-int}$ 
at zero temperature for two different hybridizations $w_c$, are shown in Fig.
 \ref{fig4}. In the upper panels, corresponding to a small $w_c$ three features
associated to the three eigenvalues of the tight-binding structure with three sites
$l=-1,0,1$, can be distinguished. Instead,
for the higher $w_c$ chosen to capture the Kondo regime used in Figs. \ref{fig2}
and \ref{fig3}, the three levels of the uncoupled structure are mixed and
only one feature can be distinguished. As a consequence of the ensuing combination of quantum
states, the transmission function as well as the current changes the sign within
a wide range of $\mu$ with respect
to the weakly coupled structure.

\begin{figure}
{\includegraphics[width=80mm, keepaspectratio,clip]{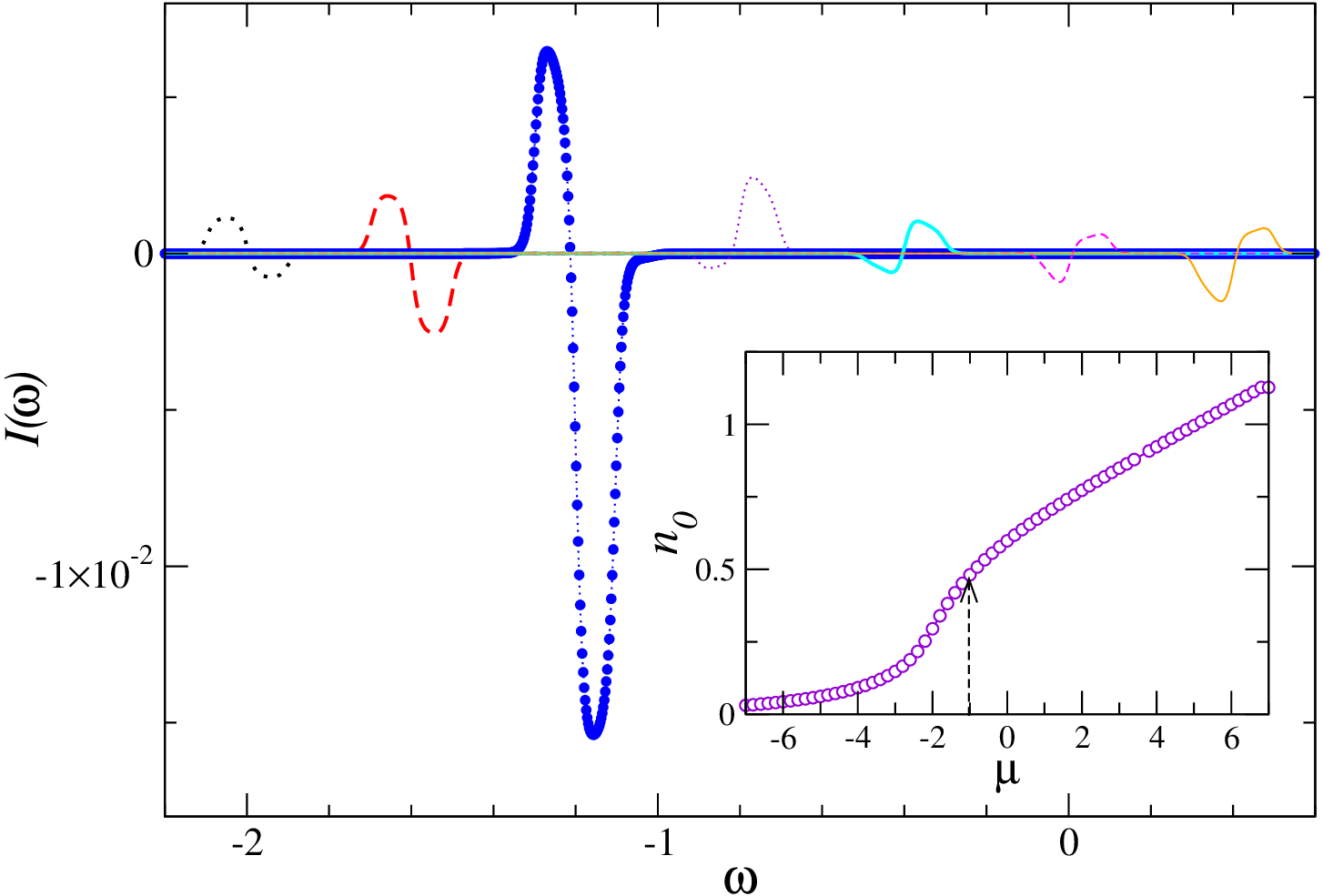}} 
\caption{(Color online) Integrand $I(\omega)$ 
for $\Omega_0=0.1$ and $U=10$. Different plots correspond to different equidistant
 values of the chemical
potential $-2 \leq \mu \leq 0.4$. In the inset, the mean occupation per spin of
the dot is shown as a function of the chemical potential.
Other parameters are as in Fig. \ref{fig2}.
}
\label{fig5}
\end{figure}
In the case of the interacting system, we analyze the behavior of the function
$I(\omega)=\sum_{k}[I^{el}(k,\omega)+I^{in}(k,\omega)]$, which, 
when integrated over $\omega$ gives the total dc current.
The behavior of this function within the Kondo regime is shown in Fig. \ref{fig5}.
Unlike the function $T(\omega)$ defined for the non-interaction system, the function
$I(\omega)$ changes as the chemical potential changes. For this reason, several plots
corresponding to different values of the chemical potential, for which
we have verified that  the Kondo resonance is developed, are shown in the figure. 
For each $\mu$ there is a feature in $I(\omega)$
around $\omega \sim \mu$ which precisely indicates the electronic transmission through the
Kondo resonance. Notice that these features are very narrow and resemble the lowest energy 
one of the upper left panel of Fig. \ref{fig4}, which corresponds to a resonance
for a dot with low hybridization with the reservoirs.
Interestingly, for $\mu \sim - 1$, $I(\omega)$ 
experiments a phase shift of $\pi$, leading to a concomitant change of sign in the dc 
current. This particular value of the chemical potential corresponds to a charge
population per spin of the dot $n_0 \equiv n^{SOSE}_{0,\sigma}(0) \sim 1/2 $.

\begin{figure}
{\includegraphics[width=80mm, keepaspectratio,clip]{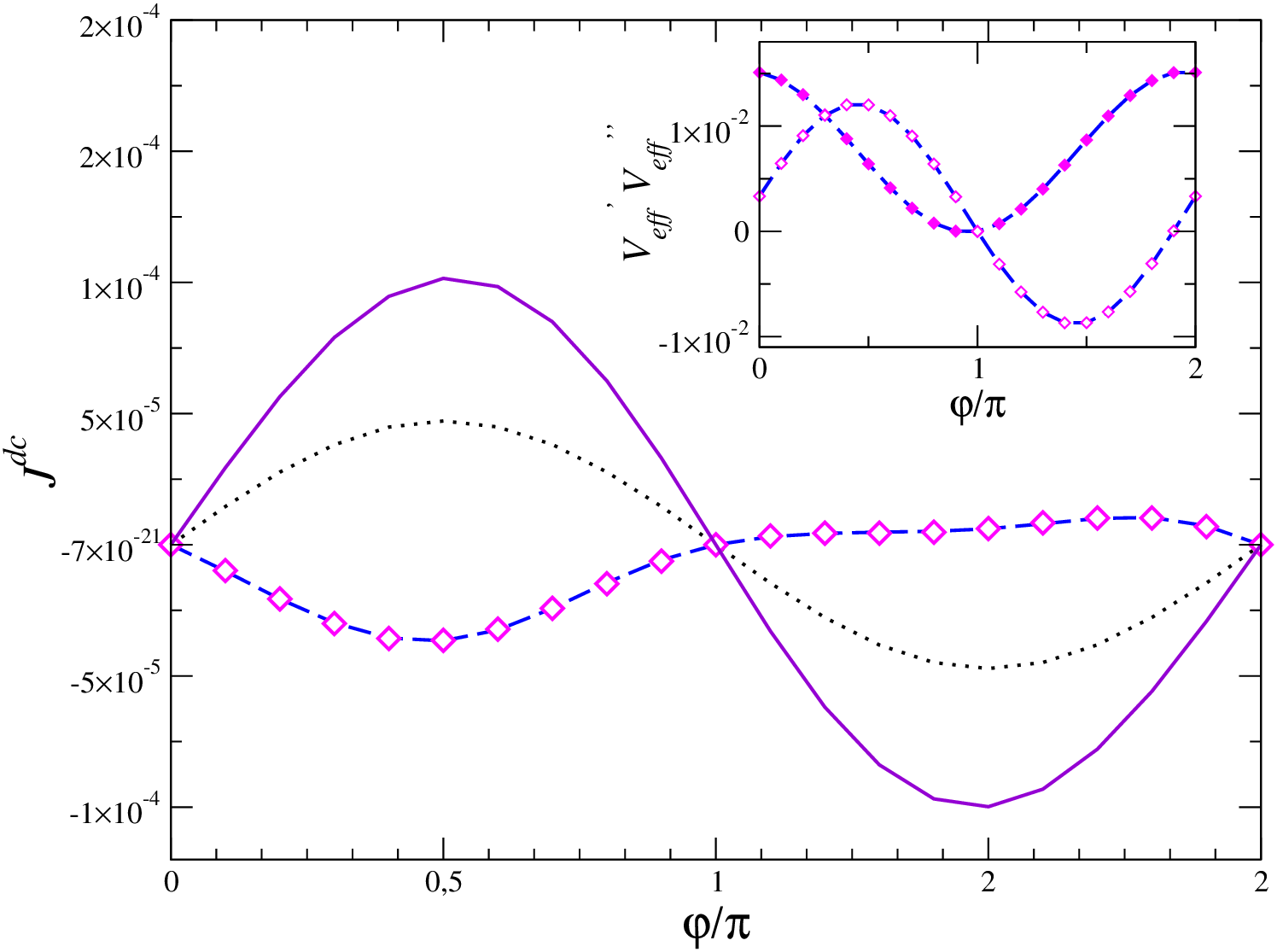}} 
\caption{(Color online) $J^{dc}$ for $\Omega_0=0.1$
as a function
of the phase-lag $\varphi$ for $U=1$ (thick solid line), and
$U=10$ (thick  dashed 
 line). The current corresponding to
the non-interacting dot (with $U=0$) is shown in dotted line, for comparison.
The symbols are fits with a function
$A_0 \sin(\varphi)+A_1\sin^2(\varphi)+ A_2  \sin(\varphi)(1+\cos(\varphi))$,
suggested by eq. (\ref{pertel}). 
Inset:  Effective potential  $V_{eff}=V_{eff}'+ i V_{eff}''$
along with fits in symbols with the function:
$A_0 \sin(\varphi)+ A_1 (1+\cos(\varphi))$. Other details are as in Fig. \ref{fig2}.
}
\label{fig6}
\end{figure}

We show in Fig. \ref{fig6} the behavior of the dc-current as a function of the phase-lag.
The limiting case of $U=0$ is plotted in dotted lines
in Fig. \ref{fig6}.
For low Coulomb interaction (see plot in solid thick line) the
induced effective pumping potential is small and the behavior of the dc-current
is qualitatively
the same as the one corresponding to the non-interacting case.
For increasing $U$, the induced pumping amplitude $V_{eff}$ becomes sizable.
Within the self-consistent procedure, this is a complex function of $\varphi$
with real and imaginary parts $V^{'}_{eff}$ and $V^{''}_{eff}$ displaying the
functional structure of $\varphi$ suggested by the perturbative solution
leading to Eq. (\ref{pertel}). The ensuing curve $J^{dc} \; \mbox{vs}  \; \varphi$
also shows the pattern predicted by  Eq. (\ref{pertel}). Notice that 
  the dc-current, as well as 
 $V_{eff}$ can be fitted with an excellent degree of accuracy
 by functions of $\varphi$ with the structure suggested by this equation 
(see symbols in  Fig. \ref{fig6}).
An striking feature observed in this figure as well as in the analytical
expression (\ref{pertel}) is the breaking of the symmetry 
$\varphi \rightarrow - \varphi$ in the behavior of the dc-current. On general
physical grounds, the symmetry of the problem in the case of identical pumping
amplitudes at the two barriers indicates that 
$J^{dc}(\varphi) \rightarrow -J^{dc}(-\varphi)$. However, the many-body treatment 
adopted in the present work breaks such symmetry for high enough induced $V_{eff}$, even
at the level of the simple self-consistent
 TDHF approximation, i.e. even disregarding self-energy 
effects. The self-consistent evaluation of higher harmonics 
(we recall that we truncate at $|k|=2$), and additional 
self-energy and vertex corrections that we have not taken into account are
expected to restore such symmetry. In any case, our results should be 
interpreted as a piece of evidence on the
departure from the $J^{dc} \propto \sin(\varphi)$ behavior induced 
by the interactions.

\begin{figure}
{\includegraphics[width=80mm,
  keepaspectratio,
clip]{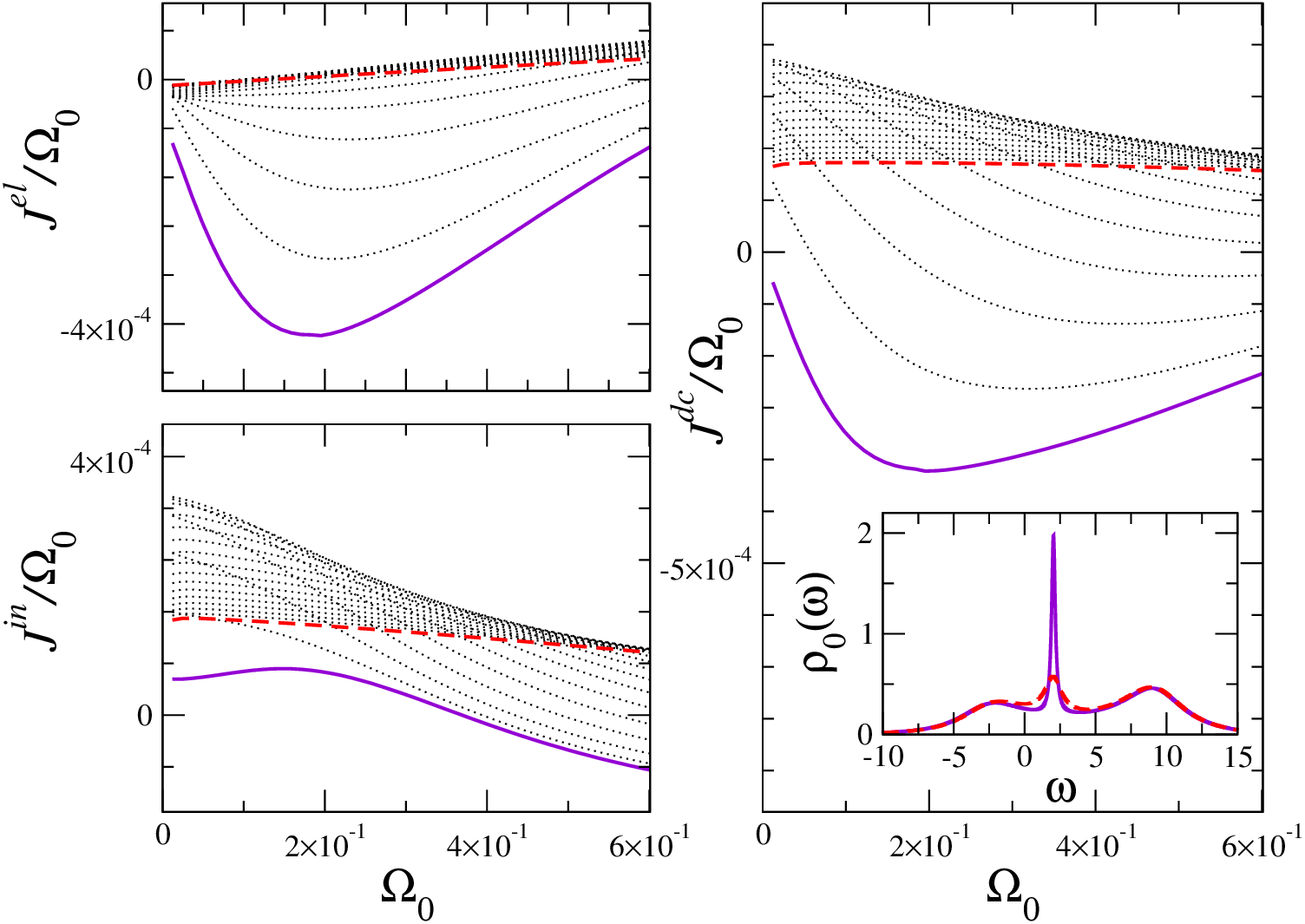}} \caption{(Color online) $J^{el} / \Omega_0$ (top left),
 $J^{in}/\Omega_0$ (bottom left) and $J^{dc}/ \Omega_0$ (right) as functions
of the pumping frequency $\Omega_0$ for $20$ values of equally spaced temperatures in the range
$0.04 \leq T/T_K \leq 0.8$ and Coulomb
interaction $U=10$. The lowest and
highest temperatures are indicated in thick solid and dashed lines, respectively. The local density
of states of the dot, $\rho_0(\omega)$,
 is depicted in the inset for the lowest and highest
temperatures. Other details are as in Fig. \ref{fig2}}
\label{fig7}
\end{figure}
  
So far, we have focused in the case of temperature $T=0$, where inelastic effects
play an insignificant role at low $\Omega_0$. To finalize, we analyze in what follows
inelastic effects, which become relevant at all frequencies at finite $T$.
The evolution of the dc-current 
 as a function of $\Omega_0$ as  the temperature grows  within a range $0<T<T_K$
is analyzed in Fig. \ref{fig7}.  At first
glance, it becomes apparent how the dominant contribution at low $\Omega_0$
 is $J^{el}$ at the lowest 
$T$, while it turns to be $J^{in}$  for the highest ones. 
The elastic contribution vanishes as the Kondo resonance disappears while the
inelastic one increases due to the corresponding grow of $\Gamma^{in}$. 
The latter increment translates into the inverse time $\tau^{-1}$, thus increasing
the range of  $\Omega_0$ where the behavior $J^{dc} \propto \Omega_0$ holds. 
In fact,
notice that the range of low $\Omega_0$ where the  plots of Fig. \ref{fig4} look
horizontal increases with increasing $T$.

\section{Summary and conclusions}

To conclude, we have analyzed a simple model for an interacting quantum
pump  by means of non-equilibrium 
Green's function techniques and within a second order self-energy 
approximation.  We have shown that
the effective time-dependent scattering center induced by
the interactions generates interference effects, which
should be detected in a $J^{dc}  \; \mbox{vs} \; \varphi$ experimental curve, following
the pattern predicted by Eq. (\ref{pertel}).
We have shown that the Kondo
effect  manifests itself in 
the $J^{dc}  \; \mbox{vs} \; \Omega_0$
 behavior,  which could be also detected in future experiments. 
Below the Kondo temperature, $T_K$, the Kondo resonance enables
the elastic transport of electrons,  however the frequency range within which $J^{dc}$ behaves 
linear in $\Omega_0$ is extremely narrow. As the temperature grows,
 inelastic scattering becomes dominant and this range  becomes wider.

\section{Acknowledgments}

We thank C. Urbina and
J. Splettstoesser for constructive comments. We
 acknowledge support from CONICET, Argentina,
 FIS2006-08533-C03-02, 
the ``RyC'' program from MCEyC, grant DGA for Groups of Excelence
 of Spain and
the hospitality of Boston University (LA), as well as FIS2005-06255 from MCEyC 
Spain (ALY and AMR).


\begin{thebibliography}{99}
\bibitem{switkes} M. Switkes,  C. M. Marcus, K. Campman, A. C. Gossard,
   Science {\bf 283}, 1905 (1999).

\bibitem{pumpex}   L. J. Geerligs, 
V. F. Anderegg, P. A. M. Holweg, J. E. Mooij, H. Pothier, D. Esteve, C. Urbina and M. H. Devoret,
   Phys. Rev. Lett. {\bf 64}, 2691 (1990); L. DiCarlo,  C. M. Marcus and J. S. Harris,
    Phys. Rev. Lett. {\bf 91}, 246804 (2003).

\bibitem{pepper}M. D. Blumenthal,   B. Kaestner, L. Li, S. Giblin, T. J. B. M.
 Hanssen, M. Pepper, D. Anderson, G. Jones and D. A. Ritchie, 
Nature Physics
{\bf 3}, 343 (2007).

\bibitem{kondodot1}D. Goldhaber-Gordon {\em et al}, Nature (London) {\bf 391}, 156 (1998).

\bibitem{kondodot2}B. Babic, T. Kontos and C. Schonenberger, 
 Physical Review B {\bf 70}, 235419, (2004); P. Jarillo-Herrero, {\em et al},
 Nature {\bf 434}, 484(2005).

\bibitem{supjunc} P.M. Billangeon, F. Pierre, H. Bouchiat, and R. Deblock,
Phys. Rev. Lett. {\bf 98}, 126802 (2007); 
S. Russo, J. Tobiska, T. M. Klapwijk, and A. F. Morpurgo, Phys. Rev. Lett. 
{\bf 99}, 086601 (2007).

\bibitem{butad}M. B\"uttiker, H. Thomas, A. Pr\^etre, Z. Phys. B {\bf 94}, 133 (1994).

\bibitem{brouwer} P. W. Brouwer, Phys. Rev. B {\bf 58}, R10135 (1998).

\bibitem{brou2001} P.W. Brouwer, Phys. Rev. B {\bf 63}, 121303 (2001);
 M.L. Polianski and P.W. Brouwer, Phys. Rev. B {\bf 64}, 075304 (2001).

\bibitem{mobu}M. Moskalets and M. B\"uttiker, Phys. Rev. B {\bf 66}, 205320 (2002);
 Phys. Rev. B {\bf 69}, 205316 (2004).  

\bibitem{otros}I. L. Aleiner and A. V. Andreev, Phys. Rev. Lett. {\bf  81}, 1286 (1998);
 F. Zhou, B. Spivak, and B. Altshuler, Phys. Rev. Lett. {\bf 82}, 608 (1999);
 J. E. Avron, A. Elgart, G. M. Graf, and L. Sadun, Phys. Rev. B {\bf 62}, R10618 (2000);
Phys. Rev. Lett. {\bf 87}, 236601 (2001); 
J. Math. Phys. {\bf 43}, 3415 (2002). 
 O. Entin-Wohlman, A. Aharony, and Y. Levinson, Phys. Rev. B {\bf 65}, 195411 (2002);
 V. Kashcheyevs, A. Aharony, and O. Entin-Wohlman,
Phys. Rev. B {\bf 69}, 195301 (2004).

\bibitem{konpump}B. Wang and J. Wang,
Phys. Rev. B {\bf 65}, 233315 (2002);
M. N. Kiselev, K. Kikoin, R. I. Shekhter, and V. M. Vinokur, 
Phys. Rev. B {\bf 74}, 233403 (2006);
E. Sela and Y. Oreg, Phys. Rev. Lett. {\bf 96}, 166802 (2006).

\bibitem{slavebos}T. Aono, Phys. Rev. Lett. {\bf 93}, 116601 (2004);
J. Splettstoesser, M. Governale, J. K\"onig and R. Fazio,
Phys. Rev. Lett. {\bf 95}, 246803 (2005).

\bibitem{das}K. Das, cond-mat/0710.2953.

\bibitem{aa}A. Levy Yeyati,  A. Martín-Rodero, and F. Flores,
 Phys. Rev. Lett. 
{\bf 71}, 2991 (1993).


\bibitem{lilipum} L. Arrachea,
   Phys. Rev. B {\bf 72}, 125349 (2005);  Phys. Rev. B {\bf 75}, 035319 (2007).

\bibitem{self} A. Oguri,
Phys. Rev. B {\bf 52}, 16727 (1995); R. L\'opez, R. Aguado, G. Platero, and C. Tejedor
Phys. Rev. Lett. {\bf 81}, 4688 (1998); A. Levy Yeyati, F. Flores, and A. 
Mart\'{\i}n-Rodero
Phys. Rev. Lett. {\bf 83}, 600 (1999).

\bibitem{lilipum1}L. Arrachea and M. Moskalets	
Phys. Rev. B {\bf 74}, 245322 (2006).	

\bibitem{lilidir} L. Arrachea, C. Na\'on, and M. Salvay,
Phys. Rev. B {\bf 76}, 165401 (2007).
 



\end{thebibliography}
\end{document}